\documentclass[final,1p,times]{elsarticle} 
\usepackage{graphicx} 
\usepackage{amssymb} 
\usepackage{amsthm} 
\usepackage{lineno} 

\usepackage[bookmarks=true,bookmarksnumbered=true]{hyperref}              

\journal{Nuclear Physics A} 
\begin{document} 

\begin{frontmatter} 


\title{Puzzles, Progress, Prospects: pre-Summary for the \\
\href{http://www.phy.ornl.gov/QM09/}{Quark Matter 2009 Conference}}

\author{W.A.~Zajc}

\address{Department of Physics, 
Columbia University, New York, NY, 10027, USA}

\begin{abstract} 
The new millennium's qualitative advances in relativistic heavy ion
physics are in part due to, and are in part causing, a new appreciation for
quantitative rigor in both experimental and theoretical work in the field.
In these proceedings for the  conference-opening ``pre-Summary'' talk I present
an annotated guide to the figures and points made in my talk\cite{zajcQM09}.
\end{abstract} 

\end{frontmatter} 



\section{Introduction}
The field of relativistic heavy ion physics was born in 1974, at a workshop
hosted at the Bear Mountain conference center\cite{BearMountain} in New York state.
The discovery of QCD\cite{Gross:1973ju,Politzer:1973fx}
during the field's infancy
\footnote{Some liberty with strict temporal has been taken in the interest of poetic license.}
led to a precocious childhood ripe with the expectation of quark-gluon plasma formation\cite{Shuryak:1980tp}
in high energy nuclear collisions.
A not atypical adolescence followed, in which vaunting ambitions were confronted with
life's realities (most notably the programmatic restrictions and/or prohibitions on
obtaining quality reference and control data at LBL's Bevalac, BNL's AGS and CERN's SPS).
With young adulthood has come maturity, a desire to ``put away childish things'', and
a consolidation of results and discoveries that promises a long and fulfilling future.
In what follows I will present a few of life's lessons that have been learned along the way.

\section{Know Your Reference}
In its most straightforward incarnation, experimental relativistic
heavy ion physics is dedicated to the search for new phenomena not
observed in the ``baseline'' measurements of p+p nor in the
``control'' measurements of p+A collisions. As alluded to above,
this has been true for several fixed-target programs. As a case in
point, consider the ``DLS Puzzle'', which refers to an excess seen
in the invariant mass of $e^+e^-$ pairs as measured by the Di-Lepton
Spectrometer in low energy ($\sim 1\ \mathrm A \cdot \mathrm{GeV}$)
C+C and Ca+Ca collisions\cite{Porter:1997rc} \footnote{RHIC and LHC
students, please note that here energy is specified as the kinetic
energy per nucleon of the beam.}. A significant excess of pairs in
the invariant mass range $0.2 < M(e^+e^-) < 0.5 \ \mathrm{GeV}$ not
accounted for by then extant transport models was observed, and
later confirmed by the HADES collaboration\cite{:2008yh}.
Theoretical understanding of the ``excess'' followed from a
corrected understanding of the critical role the $\Delta$ Dalitz
decay and of bremsstrahlung production in p+n collisions
incorporated into a detailed transport
code\cite{Bratkovskaya:2007jk}, {\em and} as verified by comparison
to measured p+p baseline and p+d control
measurements\cite{Wilson:1997sr}. The clear lesson here is the
importance of {\em in situ} measurements in the same channels in the
same apparatus in simpler systems, and attention to the fine details
(see also following section) when incorporating same into models:
{\em Know Your Reference.}

\section{Details Matter}
The first two-pion HBT measurements at
RHIC\cite{Adler:2001zd,Adcox:2002uc} clearly demonstrated that
$R_{out}/R_{side} \sim 1$, with the ratio showing only a weak
dependence on the transverse momentum of the pair. This was in stark
contrast to predictions\cite{Pratt:1986cc,Bertsch:1988db} that a
first-order phase transition would result in $R_{out}$ being
significantly larger than $R_{side}$, a result that was expected to
persist even in the case of a relatively smooth cross-over
transition\cite{Rischke:1996em}. Calculations using ideal
hydrodynamics\cite{Heinz:2001xi} produced excellent agreement with
the single-particle transverse momentum spectra and of elliptic
flow, but failed to describe the observed values of $R_{out}$ and
$R_{side}$, leading to the ``HBT Puzzle at RHIC''.

A resolution to this issue is now at hand. Recent work by
Pratt\cite{Pratt:2008qv} has demonstrated that it is not one single
factor, but rather a ``conspiracy'' of several factors that
contributed to explaining the observed values of $R_{out}$ and
$R_{side}$. These include viscous corrections, a more sophisticated
treatment of final state interactions between the two pions, a
stiffer equation of state and (perhaps most importantly, and with significant
implications for other issues in the hydrodynamic description),
pre-thermal acceleration of the matter\cite{Vredevoogd:2008id}. The
lesson here should be clear- while it is tempting to invoke
schematic models involving exotic mechanisms to ``resolve'' the HBT
puzzle, it is necessary first to analyze completely the (so-called)
mundane effects neglected in the leading-order hydrodynamic
description: {\em Details Matter.}

\section{Control Your Control Measurement}

No single theoretical prediction of a deconfinement signal has
attracted more attention than the suggestion by Matsui and
Satz\cite{Matsui:1986dk} that the charmonium states would be
suppressed in a quark-gluon plasma due to Debye
screening\footnote{This sentence is carefully worded to avoid a
misstatement I made during my talk: The Matsui and Satz paper is not
the most referenced paper in our field; Bjorken's paper on boost
invariance\cite{Bjorken:1982qr} has significantly more references. I
am grateful to Larry McLerran both for pointing this out to me and
for doing so privately rather than publicly.}. Yet despite an
extraordinary amount of theoretical and experimental activity,
ambiguities remain\cite{Levy:2009td}. Understanding the suppression
pattern will required careful measurement of the feed-down from both
$\chi_C$'s and $B$-mesons; such efforts have begun at RHIC but have
yet to reach quantitative precision. At energies above those of the
SPS the effects of $c\mbox{-}\bar{c}$
coalescence\cite{BraunMunzinger:2000px,Thews:2000rj} must be
considered, which in turn requires good knowledge of the $p_T$ and
rapidity dependence of open charm production. Cold nuclear matter
(CNM) have been carefully reconsidered at SPS energies and found
larger than previously estimated, particularly when including the
interplay with (anti)-shadowing\cite{Scomparin:2009tg}. At RHIC, CNM
effects are essentially unconstrained by the statistical precision
of the Run-3 d+Au data set, in that the errors on the break-up cross
section are comparable to its magnitude, which in turn implies that
one can not rule out the entire suppression pattern observed in
Au+Au collisions as being consistent with presence of CNM-effects
only. While all of these issues are being or will be resolved by
analysis of data sets with much greater integrated luminosity, two
lessons seem inescapable: 1)~The J$/\Psi$ is a hadron with
appreciable cross-section in nuclear matter and (even more
importantly) 2)~{\em Control Your Control Measurement}.

\section{The Shape of Things to Come}
The title of this section is taken from H.G.~Wells' evocative novel
of the same name\cite{Wells}, but both in reading Wells' work and
the remainder of this article, one should keep in mind Bohr's famous
dictum ``Prediction is very difficult, especially about the
future.''\cite{Bohr}. Therefore, in what follows work that postdates
the presentation of my talk will be cited where useful.

\subsection{Event Morphology}
Five years ago Miklos Gyulassy's call for
10-dimensional\footnote{Recent converts to the field please note
that these dimensions are {\em not} those of Anti-de Sitter space.}
tomography of RHIC collisions\cite{Gyulassy:2004vg} seemed at best
fanciful. Recent work has brought us much nearer to that goal,
although here I will focus more in morphology, that is, the study of
event shapes, rather than tomography. The outstanding example of
this is the ``ridge'', that is, the enhancement in particle
production seen in nucleus+nucleus collisions on the near-side of a
``trigger'' particle. This feature extends at least for $\Delta\eta
\sim \pm 4$ units of pseudo-rapidity about the ``trigger'' particle.
Here the quotes on ``trigger'' are used to highlight the ambiguity
(or ubiquity) of ridge phenomena\cite{ZajcRidge}. Such enhancements
have been observed in {\em untriggered} analyses\cite{Adams:2004pa},
in events with moderately high triggers ($p_T^{Trig} >
2.5$~GeV/c)\cite{Alver:2009id}, and in events with very high $p_T$
triggers\cite{Abelev:2009qa}. Note that in the latter two cases, the
ridge enhancement, while clearly present, is at most a few percent
enhancement in yield over that the bulk soft particle production.
Perhaps the most impressive morphological feature of the ridge is
its extent over a huge range in relative pseudo-rapidity. Data on
the relative extent in {\em rapidity} would be most welcome- it seems
likely that the observed correlations are indeed in relative
angle rather than in relative momentum. This, along with the fact
that ridge enhancements are seen for all values of $p_T^{trig}$ has
proven to be a very challenging to theoretical models; it is safe to
say that no model is able to incorporate the ridge over the complete
dynamic range\cite{Nagle:2009wr}.

\subsection{``Details Matter'' {\em Applied}}
The realization that thermal radiation from a quark-gluon plasma is
a key diagnostic actually predates the coinage of that name for
deconfined quarks and gluons\cite{Feinberg:1976ua,Shuryak:1978ij}.
But unambiguous detection of thermal photons, whether real or
virtual, has proved maddeningly elusive. In the case of real
photons, the difficulties are  in large part due to the (nearly)
overwhelming background of decay photons from $\pi^0$'s and $\eta$'s.
Virtual pairs evade this background, but have subtle and demanding
experimental challenges of their own, in which details truly do
matter in determining and removing all sources of background, from
both physics channels (internal conversions, correlated open charm
production) and backgrounds (external conversions, decays). Both
PHENIX\cite{:2008fqa} and NA60\cite{Arnaldi:2008er,Arnaldi:2008gp}
have met those challenges, providing first data on thermal pairs
from the medium in $\sqrt{s_{NN}}=200$~GeV Au+Au collisions and
$\sqrt{s_{NN}}=17$~GeV In+In collisions, respectively. These results
open a window on a critical new observable, which will need to be
thoroughly explored in p+p and d+A collisions as well as in
theoretical models.

\subsection{What's Past Is Prologue}
%
%
%
%
%
%
%
Just as the detection of relatively low transverse momentum thermal
photons in relativistic heavy ion collisions has been a long-sought
but recently realized goal, so has been the
observation\cite{Adler:2005ig} of higher momentum, perturbatively
produced direct photons. In principle, this would seem to satisfy a
desideratum famously expressed by Wilczek\cite{Wilczek:2000ih} in a
{\em Physics Today} article published in 2000:

\begin{quotation}
At the first level, one might hope to observe phenomena that are
very difficult to interpret from a hadronic perspective but have a
simple qualitative explanation based on quarks and gluons\ldots But
there is a second, more rigorous level that remains a challenge for
the future. Using fundamental aspects of QCD theory \ldots one can
make quantitative predictions for the emission of various kinds of
"hard" radiation from a quark-gluon plasma. We will not have done
justice to the concept of a weakly interacting plasma of quarks and
gluons until some of these predictions are confirmed by experiment.
\end{quotation}

It is striking to consider the evolution of our thinking from that
time. It is indeed true that the production rate of direct photons
is consistent with pQCD calculations, providing both a verification
of the ``binary collision counting'' paradigm and an important
baseline establishing the validity of the jet quenching results. But
at the same time these very phenomena are not interpreted as
establishing the presence of a ``weakly interacting plasma'', but
rather are important elements in establishing the {\em
strongly-coupled} quark-gluon plasma (sQGP).

There is an important third, potentially even more rigorous level in
Wilczek's hierarchy of observables: differential measurements that
extract quantitative results for transport coefficients or medium
modifications. An excellent example- ``tagged'' photon-hadron
correlations to determine medium modifications to the fragmentation
functions\cite{Wang:1996yh}- was proposed several years before
Wilczek's {\em Physics Today} article. The purity of the method,
which relies on the transparency of the medium to the photon to
establish the partonic $Q^2$, is offset by the rate
penalty\footnote{As an example, RHIC running at 5 times design
luminosity produces about one 15 GeV photon per hour in the PHENIX
central arms, which must be combined with a penalty factor of $10^{-(2-3)}$ for
measuring the complete range of the fragmentation function.} exacted
by the additional power of $\alpha_{EM}$. Only now are first results
becoming available from measurements using this
technique\cite{Adare:2009vd}, but it promises to be an important
quantitative tool for future RHIC measurements and an essential
element of the LHC program.

\subsection{Towards True Jet Quenching}
It has become common practice to refer to the discoveries of strong suppression
of high $p_T$ particles at RHIC energies\cite{Adcox:2001jp} and the
disappearance of the away-side ``jet''\cite{Adler:2002tq} as ``jet quenching'', a term coined by
Gyulassy and Plumer\cite{Gyulassy:1990ye} with due reference to Bjorken's
numerically incorrect but nonetheless seminal initial work on the topic\cite{Bjorken:1982tu}.
Of course the leading particles used to date to study suppression are not jets; they are used as a stand-in for same
due to the difficulties in detecting jets (especially with traditional jet algorithms)
in an environment with a high multiplicity of soft particles.
These difficulties arise because even statistically unlikely Poisson fluctuations in the emission
pattern of soft and moderate $p_T$ hadrons that can mimic a jet's localized
energy production are more probable than the probability of a high-$p_T$ scatter.
This issue has been understood for some time\cite{ZajcColloquium};
for example the fine segmentation
of the PHENIX calorimeter $\Delta\eta \times \Delta\phi \sim 0.01 \times 0.01$
was driven by the requirement of isolating
high-$p_T$ leading particles from the soft background.

Recently, progress has been towards true jet reconstruction at RHIC.
This has been made possible in part simply by higher integrated luminosities providing
access to higher energy jets- the soft background fluctuations have a limited reach in
fake jet energy (making this effect almost irrelevant at LHC energies). But there have
also been new developments in algorithms specifically designed for jet reconstruction
in the heavy ion environment\cite{Lai:2008zp}. Comparing results presented at this conference from two
rather different approaches to the problem\cite{Salur:2009vz,Lai:2009zq} should be very
useful for understanding any remaining systematic uncertainties in extracting
the jet signal from the background, and for then determining jet modifications
in deconfined nuclear matter.

\subsection{Quantifying Jet Quenching}
The difficult task of measuring the modification of jets and/or high-$p_T$ hadron production
in nuclear collisions is motivated by the prospect of extracting quantitative
information on the transport properties of the medium. This possibility in turn
follows from the ability to calculate the primary production rate using perturbative QCD (pQCD),
{\em and} from pQCD calculations of energy loss in the medium, typically assumed to be
homogenous or following a simply parameterized 1-d expansion. Comparison with
actual heavy ion data requires a much more sophisticated treatment of the time development.
Recently, a systematic study of three models of jet quenching has been performed
by embedding their microscopic prescriptions
in a realistic 3-d hydrodynamic model of the expanding medium\cite{Bass:2008rv}.
A gratifying level of agreement is achieved by all of the models to the single-particle
suppression measure $R_{AA}(p_T)$, at the expense of a frustratingly large range
\footnote{A factor of $\sim 4$ across these three models; inclusion of other models would
extend this range to essentially a factor of 10.} in the
associated values of the quenching parameter $\hat{q}$.
To some extent this is an illustration of the ``fragility'' of
single-observables\cite{Eskola:2004cr}, borne out by the predicted
differences of the same three models in computing more differential observables
such as the dependence of the suppression with respect to the reaction plane.
However, equally important is a consistent treatment of effects such as
collisional as well as radiative terms, assumed value of the coupling `constant',
kinematic cut-offs, assumed scale hierarchies, etc.\cite{ColeQM08,Jeon:2009yv}.
Both the work noted above and the efforts of the TECHQM\cite{jetTECHQM} collaboration have been
instrumental in highlighting these issues and in enabling their resolution.

\subsection{Hydrodynamics for the New Millennium}
The seemingly staid topic of relativistic fluid dynamics has been revolutionized in the past decade,
driven in large part by the need to describe the near-perfect fluidity observed at RHIC.
In particular, the desire to compare realistic hydrodynamic calculations with non-zero viscosity
to the experimental data necessitated a complete re-examination of the formalism.
While it has been known for some time that limiting the expansion to first-order in the gradients
led to severe instabilities\cite{Hiscock:1985zz}, the various second-order expansions were known
to be stable and thought to be complete. This has turned out not be the case; previous workers
had made assumptions appropriate for astrophysical applications that led to the dropping of
terms that are important in a completely relativistic system such as the RHIC fluid.
Valuable guidance in this exercise was obtained from the Ads/CFT correspondence, at least
for the case of conformal fluids\cite{Baier:2007ix}.

The end result of this effort, combined with extensive work on numerical implementation of the
resulting equations, has been a ``concordance'' in modeling RHIC collisions with second-order
dissipative hydrodynamics\cite{bulkTECHQM}. Impressive agreement with a wide variety of bulk observables
has been obtained\cite{Romatschke:2007mq,Song:2007ux,Luzum:2009sb}, leading to the exciting prospect
of extracting numerical values of transport coefficients. A more difficult task will be
determining the systematic uncertainties in such extractions due to ill-determined
initial conditions, eccentricity fluctuations and hadronic rescattering, but the
rate of recent progress supports cautious optimism that these challenges will be met
in the near future.

\subsection{Knowing Knudsen}
While each transport coefficient provides insight into the medium,
the ratio of shear viscosity $\eta$ to entropy density $s$ has attracted
far and away the most attention, in very large part due to the bold
conjecture\cite{Kovtun:2004de} that this ratio satisfies a ``quantum bound'' $\eta/s \ge 1/4\pi$ (in natural units).
Both the detailed hydrodynamics simulations mentioned above
and a variety of heuristic estimates\cite{Gavin:2006xd,Lacey:2006bc,Adare:2006nq,Drescher:2007cd}
suggest that $\eta/s \sim (1\mbox{-}4)/4\pi$ for the RHIC fluid, making it the most perfect
of all imperfect fluids studied in the laboratory. This provocative statement, while supported
by arguments by Danielewicz and Gyulassy\cite{Danielewicz:1984ww} which anticipated the bound on $\eta/s$,
clearly demands strong scrutiny in order to establish its validity.

An intriguing recent development has been the parametrization of
deviations in the elliptic flow pattern from that of ideal hydrodynamics
in terms of the {\em Knudsen number} $K\equiv \lambda/R$, where
$\lambda$ is the mean free path and $R$ is a characteristic system size\cite{Drescher:2007cd}.
Specifically, the centrality dependence of the elliptic flow parameter $v_2$ scaled by
the (estimated) initial state eccentricity $\epsilon$ is assumed to have the simplest
linear form consistent with ideal hydrodynamics in the limit $K \rightarrow 0$.
A fit to the experimental data produces a value for $\lambda$, which is then
used to determine the viscosity under the assumption of Boltzmann transport.
The resulting values for $\eta/s$ span a range of $0\mbox{-}10$ times the quantum bound\cite{Tang:2009vp},
a range driven in large part from extreme sensitivity to the assumptions of the ill-determined
initial state eccentricity.

There is an interesting dialectic at play here. While it is tempting to identify
the striking pattern of constituent-quark scaling\cite{Adler:2003kt,Adams:2003am} as evidence
for underlying quasi-particles, the existence of long-lived quasi-particles
is inconsistent with a strongly coupled medium near the quantum bound\cite{LindenLevy:2007gd},
this result being well-known in condensed matter physics\cite{Sachdev:2008ba}.
Conversely, hydrodynamics calculations, which correctly forgo particle transport
during the (nearly perfect) fluid phase, often suffer from failing to treat the true
particles present in the hadronic phase. Somewhat surprisingly, the determinations of $\eta/s$ based on
Knudsen number parameterizations, with their implicit assumption of ballistic transport
of quasi-particles, are not vastly different from those as determined from
second-order relativistic hydrodynamic calculations incorporating viscous effects\cite{Masui:2009pw}.
This may be viewed as a first step towards quantifying the many systematic uncertainties\cite{nszPaper}
in the Knudsen number approach, but there remains more work to be done. As but one example,
it will be important to know how realistic it is to identify a single value for $\eta/s$
to represent the time evolution of this key quantity in the course of a heavy ion collision.

\subsection{Critical Developments}
There is general agreement that the predicted first-order phase transition at high net baryon densities
between hadronic and deconfined matter should end at a critical point\cite{Stephanov:1998dy}.
This conclusion is dependent on a proper treatment using the correct $u$, $d$ and $s$ physical quark masses,
and there is still very considerable uncertainty
in the location of the critical point in the $(T,\mu_B)$ plane\cite{Fodor:2009ax}.
Similarly, while it is expected that there will be ``strong'' fluctuations in the
``vicinity'' of the critical point, much work remains to sharpen these statements
to the point where hypotheses can be rejected by observation.
The situation is not dissimilar to previous stages in our field's history,
which teaches us that qualitative advances in experimental data are necessary
for progress. Ongoing analyses of SPS data, recent low-energy results from RHIC\cite{newSTARPaper} and
ambitious future plans\cite{Heuser:2009gg,Toneev:2007yu} will provide that impetus towards
extending our understanding of the QCD phase diagram.

\section{Conclusions}
Our field is suffering, in the most pleasant sense of the word, from an embarrassment of riches.
The enormous progress made in the past decade will be sustained into the future by new low-energy facilities;
by upgraded experiments, luminosity and capabilities at RHIC; and by the unprecedented increase in
center-of-mass of energy and sophisticated experiments of the LHC, making this truly is a golden era
in relativistic heavy ion physics.



\section*{Acknowledgments}
This work was supported by U.S. Department of Energy grant
DE-FG02-86ER40281.

\end{document}